\documentclass{article}
\usepackage[preprint]{spconf}
\copyrightnotice{\copyright\ IEEE 2021}
\toappear{To appear in {\it Proc.\ ICASSP2021, June 06-11, 2021, Toronto, Ontario, Canada}}
\usepackage{amsmath,graphicx}
\usepackage{url}
\usepackage{booktabs}
\usepackage{multirow}
\usepackage[numbers,sort&compress]{natbib}
\ninept
\title{Content-Aware Speaker Embeddings for Speaker Diarisation}
\name{G. Sun, D. Liu, C. Zhang, P. C. Woodland \thanks{G. Sun is funded by the Cambridge Trust}}
\address{Cambridge University Engineering Dept., Trumpington St., Cambridge, CB2 1PZ U.K.\\
\small{\texttt{\{gs534,dl567,cz277,pcw\}@eng.cam.ac.uk}}}
\begin{document}
%
\maketitle
\begin{abstract}
Recent speaker diarisation systems often convert variable length speech segments into fixed-length vector representations for speaker clustering, which are known as speaker embeddings.
In this paper, the content-aware speaker embeddings (CASE) approach is proposed, which extends the input of the speaker classifier to include not only acoustic features but also their corresponding speech content, via phone, character, and word embeddings.  
Compared to alternative methods that leverage similar information, such as multitask or adversarial training, CASE factorises automatic speech recognition (ASR) from speaker recognition to focus on modelling speaker characteristics and correlations with the corresponding content units to derive more expressive representations. CASE is evaluated for speaker re-clustering with a realistic speaker diarisation setup using the AMI meeting transcription dataset, where the content information is obtained by performing ASR based on an automatic segmentation. Experimental results showed that CASE achieved a 17.8\% relative speaker error rate reduction over conventional methods.

\end{abstract}

\begin{keywords}
content-aware speaker embedding, diarisation, d-vector, speech recognition, distributed representation
\end{keywords}
\section{Introduction}
\label{sec:intro}

Speaker diarisation, the task to find ``Who spoke when'' in a multi-speaker audio stream, is a critical component in automatic speech recognition (ASR) systems for transcriptions of conversations, meetings, or broadcast shows. A typical diarisation system can often be divided into two stages: segmenting the audio into speaker-homogeneous intervals and clustering them into groups that should correspond to the same speaker. 
 Nowadays, the speaker clustering stage often converts variable-length speech segments into fixed-length vectors, referred to as \textit{speaker embeddings}, to characterise the speaker identity in a multi-dimensional space in which the clustering can be performed. The variations present in a spoken utterance include differences in speaker identity, content and style of the utterance,  microphone, channel and noise characteristics.
 Traditional Gaussian mixture model-based speaker dependent ASR systems either normalise the differences between speakers, or jointly model them with the phonetic units \cite{CMLLR,EIGENVOICE,phone-aware}. For both speaker recognition and diarisation, the \textit{i-vector} approach models all variations together using joint factor analysis \cite{IVECTOR}.
With deep learning, neural network (NN) models are trained to discriminate the training set speakers for each frame \cite{DVECTOR,SVECTOR} or segment \cite{selfattent1,2dselfatten,DVECTOR3,DVECTOR4,DVECTOR5,XVECTOR}. Output vectors from the penultimate layer of NN models are extracted, referred to as speaker embeddings. For simplicity, all kinds of NN-based speaker embeddings \cite{DVECTOR,SVECTOR,XVECTOR,2dselfatten} are referred to as \textit{d-vectors} without any distinction in this paper. 

Although d-vectors are very effective in encoding speaker dependent acoustic features (\textit{e.g.}, vocal tract length), they include less information about content-related features (\textit{e.g.} common terms of use) compared to i-vectors \cite{whatdoesencode}, because an NN classifier can effectively reduce the variations that are less relevant to the targets in its first few layers. One solution to this problem is to train the NN model to jointly classify both speaker and phonetic units \cite{JVECTOR} in a multi-task or adversarial learning framework \cite{multitask, adversarial, jointmultiadv}, which enforces the d-vector to encode phone related information. However, it is not straightforward to extend this method to models that are trained to extract d-vectors over hundreds of frames in a segment using a statistical pooling \cite{DVECTOR5,XVECTOR} or a self-attentive structure \cite{selfattent0,selfattent1,2dselfatten,gan,gan2}, as classification of speakers and phonetic units often require very different windows of input features.

In this paper, a content-aware speaker embedding (CASE) generation scheme is proposed, which encodes content-related features into speaker embeddings by simply extending the input of the embedding extraction network to include additional information from the speech content (at the phone, character or word level). 
Depending on the task, the speech content can be obtained from either the reference or using an ASR system. Phone, character, and word embeddings are used to encapsulate content information, and are appended to the corresponding original acoustic features according to the content-to-frame alignments. Compared to other methods, CASE does not adopt any change to the training method and the model structure except for the input layers, and is generally applicable to any type of NN used for embedding extraction. Meanwhile, CASE 
can lead to more expressive speaker embeddings since it
explicitly factorises speech recognition from the speaker embedding extraction model by explicitly conditioning on the speech content.  
The proposed CASE generation scheme was evaluated on the AMI meeting corpus. Experimental results on speaker clustering for diarisation showed that the diarisation error rate (DER) was improved by a clear margin when CASE-based d-vectors were used. Moreover, CASE can be applied to end-to-end speaker diarisation \cite{e2e1,e2e2,dnc} and also to text-dependent and -independent speaker identification and verification.

\begin{figure*}[h]
    \centering
    \includegraphics[scale=0.5]{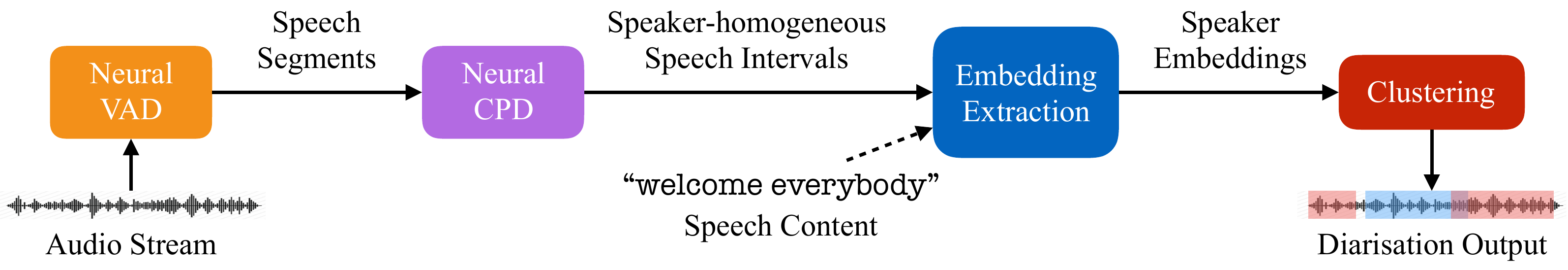}
    \caption{Our speaker diarisation pipeline. The dashed lines indicate where the content information can be added for the CASE scheme.}
    \label{fig:pipeline}
    \vspace{-0.3cm}
\end{figure*}

The remainder of this paper is organised as follows. Section~\ref{sec:2} reviews the diarisation pipeline used in this paper. Section~\ref{sec:3} presents the detailed use of CASE in  d-vector extraction. The experimental setup and results are given in Sec.~\ref{sec:4} and \ref{sec:5}, followed by the conclusion in Sec.~\ref{sec:6}.

\section{Speaker Diarisation Pipeline}
\label{sec:2}
Our speaker diarisation pipeline includes a neural voice activity detection (VAD) module, a neural change point detection (CPD) module, a speaker embedding extraction model and the clustering algorithm \cite{amidiar}. The VAD distinguishes between speech and non-speech, and selects segments corresponding to speech in the audio stream. The CPD stage splits speech segments into speaker homogeneous intervals. Then, the embedding extraction module generates d-vectors for those intervals. Finally, the spectral clustering algorithm is used to group similar intervals and one speaker label is assigned to all of them in the same group. The pipeline is illustrated in Fig \ref{fig:pipeline}.

\subsection{Neural VAD and CPD Models}
Both the VAD and CPD models are built as NN-based frame-level binary classifiers. The VAD model is a DNN of seven fully-connected layers with ReLU activation functions. The key strength of the DNN structure is the use of a large input window covering 55 consecutive frames (27 on each side), which provides sufficient information for high performance speech and non-speech classification \cite{vad}.

The CPD model adopts a ReLU recurrent neural network (RNN) model to encode past and future inputs (covering 50 frames on each side) into two vectors respectively which are then fused into one vector using the Hadamard product followed by a softmax fully-connected layer classifying the frame as a speaker change point or not.
Treating the RNN output vectors as speaker representations for the past and future audio segments, the Hadamard product and the output layer can be seen as making decisions on the change of speaker identity by comparing the speaker representations before and after the current time. 
To encourage the RNN to encode better speaker representations, frame-level d-vectors from a time-delay neural network (TDNN) model trained by classifying training set speakers for each frame are used as the input to the RNN. The CPD model including the TDNN, the RNN, and the output layer are then jointly trained to perform speaker change/non-change classification.

\subsection{Speaker Embedding Extraction and Clustering}
Speaker embeddings in this paper are the penultimate layer outputs of a TDNN-based model trained to perform classification among training set speakers. In both training and test, each variable length speech interval is first split into multiple fixed-length windows with a certain amount of overlap, and a speaker embedding is then extracted for each window. Window-level speaker embeddings, or window-level d-vectors, are extracted by aggregating frame-level d-vectors in a specific window using a multi-head self-attentive layer \cite{selfattent0,selfattent1}. In order to encode more diverse information by increasing the dissimilarities between different attention heads, a modified penalty term as described in \cite{2dselfatten} is also adopted. 

A spectral clustering algorithm \cite{rnn1} based on the cosine similarity with post-processing is used to cluster the speaker embeddings. Spectral clustering is first performed on window-level d-vectors where the number of clusters is determined by the maximum eigenvalue drop-off \cite{speccluster}. Then, variable-length speech intervals from the CPD are considered as speaker-homogeneous, and each segment is assigned to a cluster whose centroid has the smallest cosine distance to the average of the window-level d-vectors of this segment. 

\section{Content-Aware Representations}
\label{sec:3}

\subsection{Extended inputs using content}
\label{sec:3.1}
The CASE generation scheme is a method to improve the speaker embeddings by extending the input of any NN model for speaker embedding extraction to incorporate the content information corresponding to each frame. Specifically, for a model taking inputs from multiple frames, there is
\begin{align}
\vspace{-0.3cm}
	\label{eq:3}
	\begin{split}
&P({s}|\mathbf{x}_1,\ldots,\mathbf{x}_T)\\
=&\sum\nolimits_{W^{'}}P({s}|\mathbf{x}_1,\ldots,\mathbf{x}_T,W^{'})P(W^{'}|\mathbf{x}_1,\ldots,\mathbf{x}_T),
\vspace{-0.3cm}
\end{split}
\end{align}
where ${s}$ is a speaker, $W^{'}$ is a string corresponding to the content of the $T$ frames, and $\mathbf{x}_t$ represents acoustic features. The sum over all possible hypotheses often requires a lattice. To avoid intractable computation, $P(W^{'}|\mathbf{x}_1,\ldots,\mathbf{x}_T)$ is approximated by the \textit{Kronecker delta} function. Let $W$ be the reference or 1-best hypothesis string of content, $P(W^{'}|\mathbf{x}_1,\ldots,\mathbf{x}_T)$ equals to one if and only if when $W^{'}=W$ and zero otherwise. Hence Eqn.~\eqref{eq:3} can be rewritten as
\begin{align}
	\label{eq:4}
	\begin{split}
	P({s}|\mathbf{x}_1,\ldots,\mathbf{x}_T,W)=P({s}|\mathbf{x}_1,\ldots,\mathbf{x}_T,\mathbf{w}_1,\ldots,\mathbf{w}_T),
	\end{split}
\end{align}	
where $\mathbf{w}_1,\ldots,\mathbf{w}_T$ are vector representations of the time alignment of $W$. Eqn.~\eqref{eq:4} gives a simple form of CASE used throughout this paper, which extends the model input at time $t$ to be the concatenation of $\mathbf{x}_t$ and $\mathbf{w}_t$. Phone, character and word levels content information with their corresponding vector representations are explored in this paper. Phones and characters are represented using \textit{1-of-$k$} encoding which has the element at the index corresponding to the unit set to 1 and the others set to 0.  The word embeddings are derived from GloVe \cite{glove}, which represents words with compact vectors of continuous-values.

Compared to traditional methods, the CASE scheme generates more expressive embeddings by using higher-level features. Although segment-level models with joint-training on each frame can implicitly learn some of those high-level features, such learning capability depends on the length of segments whose maximum is restricted by the power of the temporal pooling method as well as the computation and storage limits (\textit{e.g.}, the GPU memory size). The CASE scheme, however, is able to incorporate more high-level features beyond the scope of a segment into the speaker embeddings, as the content sequence obtained using ASR often requires searching through a complete utterance.

Furthermore, CASE also reduces the difficulty in learning both low-level and high-level features using the same model, which can help generate more accurate speaker embeddings. It has been shown that an NN acoustic model tends to eliminate the speaker related information (such as accent or channel) using its input layers, as such information is detrimental for classifying speaker independent phonetic or linguistic units \cite{CMLLR}. Vice versa, a model trained for speaker classification does not preserve much information related to the content \cite{whatdoesencode}. This phenomenon causes a conflict in encoding both low-level and high-level (including linguistic and phonetic) features into a speaker embedding using the same set of parameters. CASE, however, addresses this issue by decoupling the linguistic and phonetic information from the speaker embedding extraction model.

\subsection{Applying CASE to d-vector extraction}
\label{sec:3.3}

The CASE generation scheme is applied to the extraction of window-level d-vectors for speaker clustering. When using CASE, it is required to obtain the content-to-frame alignments used as $\mathbf{w}_t$. 
\begin{itemize}
\item At training time, reference transcripts are first aligned with a pre-trained ASR system to derive the content (word, phone or character) vectors which are used for CASE;
\item At test time, if a manual segmentation is available, 1-best hypotheses from an ASR system can be obtained and aligned using manual segments. The aligned 1-best hypotheses will then be used as the content information in CASE. 
\item If manual segments are not available, an automatic segmentation produced by an existing speaker diarisation system will be used to get the 1-best hypotheses using ASR, which are then used to obtain the content-to-frame alignment to produce CASE-based d-vectors.

\end{itemize}
There are also situations when reference transcriptions are available for both training and test. For example, CASE can be used to improve clustering a speech corpus without speaker labels, such as in \cite{MGB}. 
The reference transcriptions can be available also in text-dependent speaker identification and verification tasks.

\section{Experimental Setup}
\label{sec:4}

\subsection{Data Preparation}
\label{sec:4.1}
All of the data preparation and model training were done using an extended version of HTK version 3.5.1 and PyHTK \cite{htk, pyhtk}. All systems were trained on the augmented multi-party interaction (AMI) meeting corpus. 
The training set contains 135 recorded meetings with 155 speakers, of which, 10\% of the data for each speaker was used for held-out validation during training. The development (\textbf{Dev}) and evaluation (\textbf{Eval}) sets from the speech recognition partition were used to evaluate the performance of proposed systems.


\subsection{Speaker Diarisation Pipeline Specification}
\label{sec:4.2}
The input acoustic features were 40-dimensional (-d) log-Mel filter banks (FBK) with a 25 ms frame size and a 10 ms frame increment, and were extracted from the multiple distance microphone (MDM) audio data pre-processed by BeamformIt \cite{beamform}. A ReLU TDNN model with 3 256-d hidden layers covering a context of [-7, +7] and a penultimate layer projecting to 128-d was used to extract frame-level d-vectors. For CPD, an RNN with 128-d hidden states was used\footnote{Note that increasing the layer width in our experimental setup did not lead to improvements in DER.}. To extract window level d-vectors, a 2-second sliding window was applied with a 1-second overlap between adjacent windows \cite{2dselfatten}. Phone and character embeddings were 48-d and 27-d respectively, and 300-d word embeddings were projected to 100-d before concatenation. The phone and character embedding layers and the word embedding projection layer were jointly trained with the speaker embedding model. Moreover, instead of using the standard softmax output activation with the cross-entropy loss, the angular softmax training loss \cite{softmax4,asoftmax,asoftmax2} was adopted with the $m$ factor set to $1$, to ensure that the derived embeddings are trained to discriminate speakers based on the cosine distance.

Model performance was evaluated in terms of the diarisation error rate (DER) which is the sum of speaker error rate (SER), missed speech (MS) and false alarm (FA). As many manual segments in AMI were found to have very long non-speech parts, which results in many unnecessary overlapping regions and also affects the clustering performance, a modified version of the manual segmentation was created by comparing each original segment with frame to speech and non-speech alignments generated by forced alignment using a pre-trained speech recognition system \cite{htk}. The {original reference} was also modified accordingly to form the {modified reference} to match the modified segmentation. Details of the original and modified segments and corresponding references, together with the speech recognition partition of the data used in this paper are available to download\footnote{\url{https://github.com/BriansIDP/AMI_diar_references.git}}.

The CASE-based d-vector systems were compared to the baseline system where d-vectors were extracted without using the content information. For completeness, SERs of two other methods that use  content information, namely multi-task training \cite{multitask} and adversarial training \cite{adversarial}, are reported. In the multi-task training method, d-vectors systems were trained with speaker and phone classification simultaneously by adding a phone classification layer. In adversarial training, a gradient reversal operation is added to the d-vector extraction layer on top of the multi-task training setup. 

\subsection{ASR System Specification}
\label{sec:4.3}
The ASR system used in this paper was trained using the Kaldi \cite{Kaldi} speech recognition toolkit based on the same HTK 40-d FBK features extracted only from the beamformed AMI MDM data. 
The speaker-independent acoustic model has six convolutional neural network layers followed by fifteen factorised TDNN blocks with residual connections, and the final layer has 2,312 output units with each of them representing a context-dependent phone. Lattice-free MMI training \cite{LFMMI} with the SpecAug data augmentation \cite{SpecAug} was used to obtain the acoustic model. A 4-gram language model estimated using the transcriptions from both AMI and Fisher corpora was used for all the decoding throughout the paper. 

\section{Experimental Results}
\label{sec:5}
\subsection{Results with manual segmentation}
The first set of diarisation experiments were performed using manual segments to directly show the effect of CASE on speaker clustering. The MS and FA rates in this case are zero, and the SER for each system is shown in Table \ref{tab:manual}. The alignments obtained using the manual segmentation and the reference transcription are denoted as \textit{reference}, while those obtained using the manual segmentation and recognition results from the ASR system is denoted as \textit{hypothesis}. 

\begin{table}[h]
    \centering
    \begin{tabular}{lcccc}
    \toprule
    \multirow{2}{*}{Systems} & \multicolumn{2}{c}{Reference (\%)} &  \multicolumn{2}{c}{Hypothesis (\%)}\\
    & Dev & Eval & Dev & Eval \\
    \midrule
    Baseline dvec. & 15.2 & 15.1 & 15.2 & 15.1\\
    Multi-task dvec. & 16.5 & 15.7 & 16.5 & 15.7 \\
    Adversarial dvec. & 12.6 & 16.7 & 12.6 & 16.7 \\
    \midrule
    CASE dvec. (p)  & 13.6 & 16.6 & 13.7	& 16.4 \\
    CASE dvec. (c) & 12.7 & 14.8 & 13.2 & 16.1 \\
    CASE dvec. (p + c)  & 11.5 & \textbf{13.3} & 11.7 & \textbf{13.4} \\
    CASE dvec. (w)  & 13.8 & 16.2 & 13.9 & 16.3 \\
    CASE dvec. (w + p + c) & \textbf{9.5} & 15.2 & \textbf{10.0} & 15.1\\
    \bottomrule
    \end{tabular}
    \caption{\%SERs with the alignment, ASR decoding, and speaker clustering all based on manual segmentation. ``w'', ``p'' and ``c'' represent the use of word, phone and character respectively.}
    \label{tab:manual}
    \vspace{-0.3cm}
\end{table}
In general, using CASE-based d-vectors consistently improves the SER on the Dev set, and the results are more varied on the Eval set, as the threshold value for spectral clustering pre-processing stage is determined based on the Dev set performance. Using the ASR hypothesis in the CASE scheme slightly degrades the SERs. Although CASE-based system with word, phone and character information achieves the lowest SER on Dev, its SER on Eval is worse than the baseline result, due to the over-fitting caused by the use of the word embeddings. 
The system with only phone and character information achieved consistently better SERs compared to the baseline on both Dev and Eval, which achieves relative SER reductions of 23\% and 11\% on Dev and Eval respectively. 
\vspace{-0.2cm}
\subsection{Results with automatic segmentation}
In this section, all speaker clustering were performed based on the automatic segments produced by the full diarisation pipeline. 
Since the use of CPD does not cause any change to MS and FA, all
SERs reported in this section can be converted to DERs by adding the MS and FA rates shown in Table \ref{tab:vad}. 
\begin{table}[h]
    \centering
    \begin{tabular}{cccc}
    \toprule
        \multicolumn{2}{c}{Dev} & \multicolumn{2}{c}{Eval} \\
        MS & FA & MS & FA \\
         \midrule
      1.2\% & 4.0\% & 1.3\% & 3.6\% \\
        \bottomrule
    \end{tabular}
    \caption{VAD MS and FA on Dev and Eval.}
    \label{tab:vad}
    \vspace{-0.3cm}
\end{table}

Speaker clustering results with the automatic segmentation are listed in Table \ref{tab:carfull}, where the hypotheses are decoded with a real ASR based on the manual segmentation. 
With CASE-based d-vectors, similar improvements to those with manual segmentation were observed, and the performance degradation caused by using hypotheses remains in a reasonable range. Our best CASE-based diarisation system achieved 10\% relative SER reductions on both Dev and Eval.

\begin{table}[h]
    \centering
    \begin{tabular}{lccccc}
    \toprule
        \multirow{2}{*}{Systems} & \multicolumn{2}{c}{Reference (\%)} &  \multicolumn{2}{c}{Hypothesis (\%)} \\
         & Dev & Eval & Dev & Eval\\
         \midrule
        Baseline dvec. &13.0 & 14.6 & 13.0 & 14.6 \\
        Multi-task dvec. & 14.3 & 13.6 & 14.3 & 13.6 \\
        Adversarial dvec. & 14.0 & 16.6 & 14.0 & 16.6 \\
         \midrule
        CASE dvec. (p) &  12.6 & 14.3  & 12.7 & 14.1 \\
        CASE dvec. (c)  & 10.8 & 14.1 & 12.1 & 14.3 \\
        CASE dvec. (p + c) & 10.7 & \textbf{12.9}  & 11.7 & \textbf{13.2} \\ 
        CASE dvec. (w) & 12.5 & 14.8  & 12.3 & 14.4  \\
        CASE dvec. (w + p + c) & \textbf{10.1} & 14.2 & \textbf{10.1} & 15.8 \\
        \bottomrule
    \end{tabular}
    \caption{\%SERs with the alignment and ASR decoding based on the manual segmentation, and the speaker clustering based on the automatic segmentation. 
    \%DERs can be calculated by adding \%FA and \%MS from Table~\ref{tab:vad} to the \%SERs.}
    \label{tab:carfull}
\end{table}

Finally, the CASE scheme is applied to the situation where neither the manual segmentation nor the reference transcripts are available, which is the same as in a real diarisation system. The hypotheses were obtained by decoding with ASR based on the automatic segmentation, which were aligned and used for the CASE-based d-vector extraction. Specifically, at test-time, the audio stream of a meeting is first passed through VAD, CPD, and the baseline d-vector extraction and clustering, to obtain the automatic segmentation. Then ASR is used to decode such automatic segments to derive the hypotheses, which are then aligned and used for CASE-based d-vector extraction and re-clustering. 
The word error rates (WERs) with both manual and automatic segmentation are listed in Table \ref{tab:wer}.

\begin{table}[h]
    \centering
    \begin{tabular}{ccccc}
    \toprule
        \multicolumn{2}{c}{Dev \%WER} & \multicolumn{2}{c}{Eval \%WER} \\
        manual & automatic & manual & automatic \\
        \midrule
        35.7 & 39.7 & 38.7 & 42.9 \\
        \bottomrule
    \end{tabular}
    \caption{\%WERs on Dev and Eval generated by our ASR using the MDM data based on the manual and automatic segmentation.}
    \label{tab:wer}
    \vspace{-0.3cm}
\end{table}

Since our current diarisation pipeline is not able to detect overlapping speech, the overlapping regions are decoded as the non-overlapping regions which generates more deletion errors and degrades the overall WERs.
Then, the CASE-based d-vectors were extracted based on the hypotheses, which is followed by the re-clustering, and the SERs are shown in Table \ref{tab:twopass}.
\begin{table}[h]
    \centering
    \begin{tabular}{lcc}
        \toprule
        Systems & Dev \%SER & Eval \%SER \\
         \midrule
        Baseline dvec. & 13.0 & 14.6 \\
        Multi-task dvec. & 14.3 & 13.6\\
        Adversarial dvec. & 14.0 & 16.6 \\
        \midrule
        CASE dvec. (p) & 13.2 & 14.3 \\
        CASE dvec. (c) &  11.8  & 15.1 \\
        CASE dvec. (p + c) & {11.0} & \textbf{12.0} \\
        CASE dvec. (w) & 13.1 & 14.7 \\
        CASE dvec. (w + p + c) & \textbf{10.8} & 16.2 \\
        \bottomrule
    \end{tabular}
    \caption{\%SERs with the alignment, ASR decoding, and speaker clustering all based on the automatic segmentation. \%DERs can be calculated by adding \%FA and \%MS from Table~\ref{tab:vad} to the \%SERs.}
    \label{tab:twopass}
\end{table}

From Table~\ref{tab:twopass}, the CASE-based systems in general outperform the baseline and the other methods tested. The best performing system is the CASE-based d-vector system with both phone and character embeddings, which achieves 15.4\% and 18.6\% relative SER reductions compared to the baseline on Dev and Eval respectively. 

\section{Conclusions}
\label{sec:6}
This paper proposes the CASE scheme which incorporates speech content information into speaker embedding extraction for speaker diarisation. The CASE scheme is applied to speaker embedding extraction stage by appending each input acoustic feature with its corresponding phone, character or word representations. Experiments on the AMI corpus used both an oracle setup with manual segmentation and reference transcriptions, and a realistic setup with automatic segmentation and hypothesis transcriptions. Our best performing CASE-based system with both phone and character embeddings consistently outperforms all baseline methods under all conditions. 


\bibliographystyle{IEEEbib}
\bibliography{strings,refs}

\begin{thebibliography}{8}

 \bibitem{vad}
L.~Wang, C.~Zhang, P.C.~Woodland, M.J.F.~Gales, P.~Karanasou, P.~Lanchantin, X.~Liu \& Y.~Qian,
\newblock { ``Improved {DNN}-based segmentation for multi-genre broadcast audio"},
 \newblock  {\em Proc. ICASSP}, Shanghai, 2016.
 
  \bibitem{selfattent0}
Z.~Lin, M.~Feng, C.N.~dos Santos, M.~Yu, B.~Xiang, B.~Zhou \& Y.~Bengio,
\newblock { ``A structured self-attentive sentence embedding"},
 \newblock  {\em Proc. ICLR}, Toulon, 2017.

  \bibitem{selfattent1}
Y.~Zhu, T.~Ko, D.~Snyder, B.~Mak \& D.~Povey,
\newblock { ``Self-attentive speaker embeddings for text-Independent speaker verification"},
 \newblock  {\em Proc. ICASSP}, Calgary, 2018.
 
   \bibitem{2dselfatten}
G.~Sun, C.~Zhang \& P.C.~Woodland,
\newblock { ``Speaker diarisation using {2D} self-attentive combination of embeddings"},
 \newblock  {\em Proc. ICASSP}, Brighton, 2019.
 
    \bibitem{rnn1}
Q.~Wang, C.~Downey, L.~Wan, P.~Mansfield \& I.L. Moreno,
\newblock { ``Speaker diarization with {LSTM}"},
 \newblock  {\em Proc. ICASSP}, 2018.

    \bibitem{speccluster}
U.~von Luxburg.
\newblock { ``A tutorial on spectral clustering"},
 \newblock  {\em Statistics and Computing}, 17, pp.395-416, 2018.

    \bibitem{glove}
J.~Pennington, R.~Socher \& C.D.~Manning,
\newblock { ``GloVe: Global vectors for word representation"},
 \newblock  {\em Proc. EMNLP}, Doha, 2014.
 
     \bibitem{whatdoesencode}
S.~Wang, Y.~Qian \& K.~Yu,
\newblock { ``What does the speaker embedding encode"},
 \newblock  {\em Proc. Interspeech}, Stockholm, 2017.
 
      \bibitem{CMLLR}
 M.~Ferras, C.C.~Leung, C.~Barras \& J.~Gauvain,
\newblock { ``Constrained MLLR for speaker recognition"},
 \newblock  {\em Proc. ICASSP}, 2007.
 
       \bibitem{MGB}
P.~Bell, M.J.F.~Gales, T.~Hain, J.~Kilgour, P.~Lanchantin, X.~Liu, A.~McParland, S.~Renals, O.~Saz, M.~Wester, P.C.~Woodland,
\newblock { ``The MGB challenge: Evaluating multi-genre broadcast media recognition"},
 \newblock  {\em Proc. ASRU}, Scottsdale, 2015.
 
        \bibitem{htk}
S.~Young, G.~Evermann, M.~Gales, T.~Hain, D.~Kershaw, X.~Liu, G.~Moore, J.~Odell, D.~Ollason, D.~Povey and A.~Ragni, V.~Valtchev, P.~Woodland \& C.~Zhang,
\newblock { ``The HTK Book (HTK 3.5)"},
 \newblock  {\em Cambridge University Engineering Department}, 2015.
 
        \bibitem{pyhtk}
C.~Zhang, F.~Kreyssig, Q.~Li \& P.C.~Woodland,
\newblock { ``{PyHTK}: {P}ython library and {ASR} pipelines for {HTK}"},
 \newblock  {\em Proc. ICASSP}, Brighton, 2019.

        \bibitem{EIGENVOICE}
P.~Kenny, G.~Boulianne \& P.~Dumouchel,
\newblock { ``Eigenvoice modeling with sparse training data"},
 \newblock  {\em IEEE Transactions on Speech and Audio Processing}, 13(3), pp.345-354, 2005.

\bibitem{IVECTOR}
S.H.~Shum, N.~Dehak, E.~Chuangsuwanich, D.~Reynolds \& J.~Glass,
\newblock { ``Exploiting intra-conversation variability for speaker diarization"},
 \newblock  {\em Proc. Interspeech}, 2011.
 
 \bibitem{XVECTOR}
D.~Snyder, D.~Garcia-Romero, G.~Sell, D.~Povey \& S.~Khudanpur,
\newblock { ``X-vectors: {R}obust {DNN} embeddings for speaker recognition"},
 \newblock  {\em Proc. ICASSP}, Calgary, 2018.

\bibitem{DVECTOR}
E.~Variani, X.~Lei, E.~McDermott, I.L.~Moreno \& J.~Gonzalez-Dominguez,
\newblock { ``Deep neural networks for small footprint text-dependent speaker verification"},
 \newblock  {\em Proc. ICASSP}, Florence, 2014.

\bibitem{SVECTOR}
Y.Z.~Işik, E.~Hakan \& S.~Ruhi,
\newblock { ``S-vector: A discriminative representation derived from i-vector for speaker verification"},
 \newblock  {\em Proc. EUSIPCO}, Nice, 2015.
 
 \bibitem{JVECTOR}
Y.~Liu, Y.~Qian, N.~Chen, T.~Fu, Y.~Zhang \& K.~Yu,
\newblock { ``Deep feature for text-dependent speaker verification"},
 \newblock  {\em Speech Communication}, 73, pp.1-13, 2015.

 
  \bibitem{DVECTOR3}
 F.A.R.R.~Chowdhury, Q.~Wang, I.L.~Moreno \& L.~Wan,
\newblock { ``Attention-based models for text-dependent speaker verificatio"},
 \newblock  {\em Proc. ICASSP}, Calgary, 2018.

  \bibitem{DVECTOR4}
G.~Heigold, I.L.~Moreno, S.~Bengio \& N.~Shazeer,
\newblock { ``End-to-end text-dependent speaker verification"},
 \newblock  {\em Proc. ICASSP}, 2016.

\bibitem{DVECTOR5}
D.~Garcia-Romero, D.~Snyder, G.~Sell, D.~Povey \& A.~McCree,
\newblock { ``Speaker diarization using deep neural network embeddings"},
 \newblock  {\em Proc. ICASSP}, New Orleans, 2017.
 
 
  \bibitem{softmax4}
W.~Liu, Y.~Wen, Z.~Yu, M.~Li, B.~Raj \& L.~Song,
\newblock { ``{D}eep hypersphere embedding for face recognition"},
 \newblock  {\em Proc. CVPR}, 2017.
 
  \bibitem{asoftmax}
Z.~Huang, S.~Wang \& K.~Yu,
\newblock { ``Angular softmax for short-duration text-independent speaker verification"},
 \newblock  {\em Proc. Interspeech}, Hyderabad, 2018.

\bibitem{asoftmax2}
Y.~Fathullah, C.~Zhang \& P.C.~Woodland,
\newblock { ``Improved large-margin softmax loss for speaker diarisation"},
 \newblock  {\em Proc. ICASSP}, Barcelona, 2020.
 
 \bibitem{e2e1}
A.~Zhang, Q.~Wang, Z.~Zhu, J.~Paisley \& C.~Wang,
\newblock { ``Fully supervised speaker diarization"},
 \newblock  {\em Proc. ICASSP}, Brighton, 2019.

 \bibitem{e2e2}
Y.~Fujita, N.~Kanda, S.~Horiguchi, K.~Nagamatsu \& S.~Watanabe,
\newblock { ``End-to-end neural speaker diarization with permutation-free objectives"},
 \newblock  {\em arXiv 1909.05952}, 2019.
 
 \bibitem{dnc}
Q.~Li, F.L.~Kreyssig, C.~Zhang \& P.C.~Woodland,
\newblock { ``Discriminative neural clustering for speaker diarisation"},
 \newblock  {\em arXiv 1910.09703}, 2019.

 \bibitem{Kaldi}
D.~Povey, A.~Ghoshal, G.~Boulianne, L.~Burget, O. Glembek, N. Goel, M. Hannemann, P. Motl$\acute{i}\check{c}$ek, Y. Qian, P. Schwarz, J. Silovsk$\acute{y}$, G. Stemmer \& K.~Vesel$\acute{y}$,
\newblock { ``The Kaldi speech recognition toolkit"},
 \newblock  {\em Proc. ASRU}, Hawaii, 2011.

  \bibitem{LFMMI}
D. Povey, V. Peddinti, D. Galvez, P. Ghahrmani, V. Manohar, X. Na, Y. Wang \& S. Khudanpur,
\newblock { ``Purely sequence-trained neural networks for ASR based on lattice-free MMI"},
 \newblock  {\em Proc. Interspeech}, San Francisco, 2016.

 \bibitem{SpecAug}
D.S. Park, W. Chan, Y. Zhang, C.-C. Chiu, B. Zoph, E.D. Cubuk \& Q.V. Le,
\newblock { ``SpecAugment: A simple data augmentation method for automatic speech recognition"},
 \newblock  {\em Proc. Interspeech}, Graz, 2019.
 
 \bibitem{beamform}
X.~Anguera, C.~Wooters \& J.~Hernando
\newblock { ``Acoustic beamforming for speaker diarization of meetings"},
 \newblock  {\em IEEE Trans. on Audio, Speech and Language Processing}, 15(7), pp. 2011--2022, 2007.
 
 \bibitem{multitask}
Y.~Liu, L.~He, J.~Liu \& M.T.~Johnson 
\newblock { ``Introducing phonetic information to speaker embedding for speaker verification"},
 \newblock  {\em EURASIP Journal on Audio, Speech, and Music Processing}, 19, 2019.

\bibitem{adversarial}
Z.~Meng, Y.~Zhao, J.~Li \& Y.~Gong
\newblock { ``Adversarial speaker verification"},
 \newblock  {\em Proc. ICASSP}, Brighton, 2019.
 
 \bibitem{jointmultiadv}
Z.~Chen, S.~Wang, Y.~Qian \& K.~Yu
\newblock { ``Channel invariant speaker embedding learning with joint multi-task and adversarial training"},
 \newblock  {\em Proc. ICASSP}, Barcelona, 2020.
 
 \bibitem{gan}
I.~Goodfellow, J.~Pouget-Abadie, M.~Mirza, B.~Xu, D.~Warde-Farley, S.~Ozair, A.~Courville \& Y.~Bengio
\newblock { ``Generative adversarial networks"},
 \newblock  {\em Proc. NIPS}, Montreal, 2014.

 \bibitem{gan2}
M.~Pal, M.~Kumar, R.~Peri, T.J.~Park, S.~H.~Kim, C.~Lord, S.~Bishop \& S.~Narayanan
\newblock { ``Meta-learning with latent space clustering in generative adversarial network for speaker diarization"},
 \newblock  {\em arXiv:2007.09635}, 2020.

 \bibitem{amidiar}
G.~Sun, C.~Zhang \& P.C.~Woodland
\newblock { ``Combination of deep speaker embeddings for diarisation"},
 \newblock  {\em arXiv:2010.12025}, 2020.
 
  \bibitem{phone-aware}
 Y.~Lei, N.~Scheffer, L.~Ferrer \& M.~McLaren
 \newblock { ``A novel scheme for speaker recognition using a phonetically-aware deep neural network"},
 \newblock  {\em Proc. ICASSP}, Florence, 2014.

\end{thebibliography}

\end{document}